# Deep forecasting of translational impact in medical research

Amy P.K. Nelson, MBChB[1*], Robert J. Gray, PhD[1], James K. Ruffle, MBBS[1], Henry C. Watkins, PhD[1], Daniel Herron, PhD[2], Nick Sorros[3], Danil Mikhailov, PhD[3], M. Jorge Cardoso, PhD[4], Sebastien Ourselin, PhD[4], Nick McNally, PhD[2], Bryan Williams, PhD[2,5], Geraint E. Rees, PhD[1,6], & Parashkev Nachev, PhD[1*].

[1]UCL Queen Square Institute of Neurology, University College London, London, UK, WC1B 5EH

[2]Research & Development, NIHR University College London Hospitals Biomedical Research Centre, London, UK, WC1E 6BT

[3]Wellcome Data Labs, Wellcome Trust, London, UK, NW1 2BE

[4] School of Biomedical Engineering & Imaging Sciences, King's College London, London, UK, WC2R 2LS

[5]UCL Institute of Cardiovascular Sciences, University College London, London, UK, WC1E 6BT

[6]Faculty of Life Sciences, University College London, Gower Street, London, UK, WC1E 6BT

*Correspondence to Amy PK Nelson (amy.nelson@ucl.ac.uk) and Parashkev Nachev (p.nachev@ucl.ac.uk).

High Dimensional Neurology Group
UCL Queen Square Institute of Neurology
Russell Square House
Bloomsbury
London, United Kingdom
WC1B 5EH

# Abstract

The value of biomedical research—a $1.7 trillion annual investment—is ultimately determined by its downstream, real-world impact. Current objective predictors of impact rest on proxy, reductive metrics of dissemination, such as paper citation rates, whose relation to real-world translation remains unquantified. Here we sought to determine the comparative predictability of future real-world translation—as indexed by inclusion in patents, guidelines or policy documents—from complex models of the abstract-level content of biomedical publications versus citations and publication meta-data alone. We develop a suite of representational and discriminative mathematical models of multi-scale publication data, quantifying predictive performance out-of-sample, ahead-of-time, across major biomedical domains, using the entire corpus of biomedical research captured by Microsoft Academic Graph from 1990 to 2019, encompassing 43.3 million papers across all domains. We show that citations are only moderately predictive of translational impact as judged by inclusion in patents, guidelines, or policy documents. By contrast, high-dimensional models of publication titles, abstracts and metadata exhibit high fidelity (AUROC > 0.9), generalise across time and thematic domain, and transfer to the task of recognizing papers of Nobel Laureates. The translational impact of a paper indexed by inclusion in patents, guidelines, or policy documents can be predicted—out-of-sample and ahead-of-time—with substantially higher fidelity from complex models of its abstract-level content than from models of publication meta-data or citation metrics. We argue that content-based models of impact are superior in performance to conventional, citation-based measures, and sustain a stronger evidence-based claim to the objective measurement of translational potential.

# Significance statement

The value of biomedical research—widely indexed by citation metrics—is ultimately determined by real-world impact whose predictability from simple indices remains unquantified. Comprehensively surveying three decades of biomedical research—43.3 million papers—we show that citations are only moderately predictive of translational impact as judged by inclusion in patents, guidelines, or policy documents. By contrast, high-dimensional models of publication titles, abstracts and metadata exhibit high fidelity (AUROC > 0.9), generalise across time and thematic domain, and transfer to the task of recognizing papers of Nobel Laureates. A revaluation of the use of quantitative metrics in guiding translation is thereby compelled: complex models offer stronger evidence for translation-focused policy, combining the objectivity and reproducibility of citations with greater predictive fidelity.

# Introduction

Scientometrics has existed for only a small fraction of the history of science itself, sparked by the logical empiricists of the Vienna Circle in their philosophical quest to construct a unified language of science(1). Developed into the familiar, citation-centred form through arduous manual extraction by Garfield and Price in the mid-20th century(2, 3), its indicators have proliferated in the Internet age. They now dominate the research landscape, routinely informing major funding decisions and academic staff recruitment worldwide(4–8).

The importance of the original goal has become magnified over time: to measure scientific progress regardless of funding or ideology, uncoloured by the reputations of individuals or institutions. But the fundamental focus of its current solution—the volume and density of discussion in print—is both detached from the ultimate, real-world objective, and subject to familiar distortions such as the popularity of papers notable only for being spectacularly wrong(9–11).

These concerns are amplified in medical science, whose primary focus is not merely knowledge, but impact on patient health: necessarily a consequence rather than a constitutive characteristic of research activity, neither easily benchmarked nor directly optimised. And there is no doubt that optimisation is needed: over the past 60 years, the number of new drug approvals per unit R&D spend has consistently halved every nine years while published medical research has doubled with the same periodicity(12), and only 0.004% of basic research findings ultimately lead to clinically useful treatments(13). The critical pre-requisite for all research—funding—shows substantial randomness in its distribution(14), enough for at least one major healthcare funder to award grants by lottery(15). Any decision function based on random chance, or a process demonstrably not much better than random chance, leaves room for improvement, particularly when commanding approximately $1.7 trillion global annual investment across the US, Japan, South Korea and the European Union(16).

Is this state of affairs partially caused by over-reliance on misleading scientometrics, have we simply not found the right metrics yet, or is the relation between scientific activity and consequent impact opaque to objective analysis? To address these crucial questions, we need a fully-inclusive survey of published medical research that relates its characteristics to an independently measured translational outcome as close to real-world impact as can be quantified. This relationship must be explored with models of sufficient expressivity to detect complex relations between many candidate predictive factors *beyond* paper-to-paper citations.

Here we provide the first comprehensive, field-wide analysis of translational impact measured by its most widely accepted proximal indices—patents, guidelines or policies—based on 30 years of data from the medical field. We quantify the ability to predict inclusion in future patents, guidelines or policies from conventional age-normalised citation counts, and compare this with the predictive fidelity of deep learning models incorporating more complex features extracted from metadata, titles and abstracts. The breadth and depth of analysis allows us to draw strong conclusions about the application of metrics to translational forecasting with substantial implications for research policy.

## Results

### Citations

Over the period from January 1990 to March 2019, only 17.1 million of the 43.3 million published papers categorised as medical by Microsoft Academic Graph were cited at least once. Of these, 964403 were included in a patent, and 16752 in a guideline or a policy document. Included papers were more frequently cited, but the numbers of citations and inclusions were weakly correlated (Pearson's r = 0.094 for guidelines or policies, r = 0.248 for patents, fig. 1).

### Predictive performance

Attempting to predict inclusion in a guideline or policy document from the traditional measure of impact—annual paper citations—yielded a mean cross-validated AUROC of 0.77 with univariable logistic regression. By contrast, a high-dimensional model trained on metadata, and title and abstract embeddings, based on a hybrid multilayer perceptron (MLP) and convolutional neural network (CNN), achieved an area under the receiver operating curve (AUROC) of 0.915 and average precision (AP) of 0.919 on unseen test data (fig. 2A, fig.

S1). The MLP trained on only metadata, without title or abstract embeddings, achieved a lower mean cross-validated AUROC of 0.88, significantly so as judged by cross-validation confidence intervals.

For the task of predicting patent inclusions, univariable logistic regression on annual paper citations yielded a mean cross-validated AUROC of 0.756. A high-dimensional model trained on metadata, and title and abstract embeddings, achieved much higher AUROC of 0.918 and AP of 0.859 on unseen test data, fig. 2B. The MLP trained on only metadata, without title or abstract embeddings, achieved a lower mean cross-validated AUROC of 0.876.

Across both tasks, a high-dimensional neural network model trained on both metadata and content embeddings substantially outperforms more commonly used paper citation based metrics when predicting future translational impact.

### Performance over time and across research domain

To test the generalisability of the models, we must examine sustained performance over time and across domains. For guideline or policy documents, high-dimensional models trained only on data from 1990 to 2013 and tested on out-of-sample papers published over the subsequent four years achieved an AUROC of 0.920 and an AP of 0.911. Crucially, there was no appreciable diminution in fidelity over time (fig. 3A&B, fig. S2A). Performance was consistently good-to-excellent within each of the top 8 most common domains of medicine (fig. 3C&D).

An identical analysis of patent inclusions produced a similar picture, yielding an AUROC of 0.902 and an AP of 0.606 for out-of-sample papers published the succeeding four years, with no diminution over time (fig. 3E&F). Indeed, the AUROC improved with time (fig. S2B). Future performance was consistently good-to-excellent within each of the top 8 most common domains (fig. 3G&H).

### Performance under metadata restriction

Rebuilding the foregoing models without paper-level metrics—paper citations, paper rank and paper mentions—and more stringently, also without attributes influenced by factors of merit extrinsic to the paper itself—affiliation, authors, journal and field—yielded slightly diminished fidelity. For guideline and policy inclusions, the combined content and metadata without paper-level metrics achieved AUROCs of 0.91, declining to 0.90 on further removal of extrinsic factors. The corresponding values for models based on metadata only were 0.83 and 0.82. A model trained only on title-abstract embeddings, without any metadata at all, achieved an AUROC of 0.89 (fig. S3A). For patent inclusions, identically constrained models yielded AUCs of 0.89, 0.87, 0.85, 0.81 and 0.82 respectively (fig. S3B).

### Transfer to Nobel Laureate paper prediction

If the high-dimensional models are capable of capturing fundamental features of translational impact, they may identify papers whose impact is judged by other criteria. To test for such *transfer learning* we applied our best patent model—*without retraining to new targets*—to the task of identifying the papers of Nobel prize-winners in Physiology or Medicine from 1991 to 2019 published before the prize was awarded. There were 166 such papers with abstracts in MAG, of which 60 were included in patents. Strikingly, the best model retrieved a higher proportion of Nobel laureate papers (103/166) than metadata (86/166) or age-normalised citations (23/166), while retaining superior fidelity for detecting patent inclusions (AUROC 0.79 vs. 0.73 and 0.73 respectively).

### Predictors of inclusion

A complex, high-dimensional model cannot easily yield intelligible weightings of predictive importance because its decision is a highly non-linear function of a large set of input features. A coarse indication of relative feature importance can nonetheless be derived from alternative architectures of lesser flexibility. Here a boosted trees (AdaBoost) model was used, trained on the metadata and optimised by gridsearch to similar performance as the MLP (AUROC 0.88, guideline or policy inclusions; AUROC 0.877, patent inclusions) (table S1).

For guideline or policy inclusions, the rank of the paper—a metric provided by Microsoft Academic Graph (MAG)(17) reflecting the eigencentrality-based 'influence' of a paper—had the highest feature importance, followed by the paper count, citation count and rank of the journal in which the paper was published. For patent inclusions, the top three features were related to journal productivity-related metrics.

### Deep semantic structure of titles and abstracts

Textual analysis—of either title or abstract—cannot easily yield an intelligible set of predictive features, as in the foregoing models. But we can visualise the sentence-level embeddings of the title and abstract encoded by BioBERT(18)—a rich, context-aware representation of natural language concepts tuned on biological text—through a succinct representation generated by a deep autoencoder. Represented in a two-dimensional space through non-linear dimensionality reduction, the embeddings showed a degree of disentanglement of clusters rich in guideline or policy inclusions vs. none (fig. 4A). This reveals intrinsic structure in the data exploited by the hybrid model to achieve the high classification performance observed. An identical analysis of the structure of the patent inclusion embeddings revealed similar intrinsic structure (fig. 4B).

### Graph community structure

The similarity and dissimilarity between papers can be modelled as a graph whose edges index the dependencies between individual features. Hierarchically arranged distinct patterns of similarity—the graph's community structure—can then be revealed by stochastic block modelling(19), here performed separately for guideline or policy-included papers, and patent-included papers, each compared against all other papers.

Distinct communities of author, institutional, journal, and domain features emerged across both groups (fig. 5 & 6). Overall, the community structures of papers included neither in guidelines or policies or patents were most similar, as indexed by pairwise comparisons of the log normalised mutual information of the inferred model parameters, and the community structure of guideline or policy-included papers was most distinctive (fig. S4). This observation cohered with the structure of an undirected features graph, weighted by the absolute correlation coefficient between features, that showed patent inclusions to be more centrally embedded within the wider network of metadata than guideline or policy inclusions (fig. S5).

Contrasting the effect of inclusion within the guideline or policy group, indices related to the first author and journal were more decisive in the included papers, whereas indices related to the institution and journal were more decisive in the others (fig. S6A). The domains of virology, endocrinology, alternative medicine, psychiatry, nursing and environmental health were also more prominent in the former, whereas surgery, radiology, traditional medicine, and rehabilitation in the latter. The effect of inclusion within the patent group was most strongly manifest in institutional indices for the included group, and field indices for the others. The contrast between domains was more striking than in the guideline or policy model, pharmacology being especially dominant in the included group, and general medical specialties in the others (fig. S6B).

### The translational impact of journals

Journal impact factors—indices of the annual citation return of an average paper—exclude patent, guideline or policy inclusions. So ranked, the top 10 journals in the medical domain based on cited papers published between 1990 and 2019 are listed in Table 1A. This corresponds to a medical domain 'impact factor' over three decades, rather than the commonly reported annual. The equivalent ranking for guideline and patent inclusions, identically filtered, are listed in Tables 1B and 1C respectively. Note that in the absence of plausibly objective weighting of policies or guidelines, this metric will be sensitive to the numerosity of distinct policy documents within any given domain, reflecting its political, regulatory, or administrative complexity.

### Discussion

We provide the first comprehensive framework for forecasting the translation of published medical research in the form of patent, guideline or policy inclusions, revealing the community structure of translational inclusions, and compute the top translationally relevant journals across biomedicine over the past 30 years.

We show that standard citation metrics are markedly inferior to those derived from complex models based on more detailed descriptions of published research. *If* objective metrics are to be used in translational assessment, then the use of conventional metrics is here shown to be insupportable.

Our analysis suggests the problem rests not with citations, but with the expressivity of any simple metric of something as constitutionally complex as research translation. It is clear that the translational signal is distributed through the combinatorial fabric of paper citation networks, metadata, and content captured in titles and abstracts. No easily interpretable scalar value could capture it. Conversely, that surprisingly economical information about a paper—its metadata, title and abstract—can be exploited by the right modelling architecture to yield high predictive fidelity, means no one could argue that no objective alternative is available. Even without full text information, we can confidently identify large swathes of research activity unlikely to inform guidelines or policy, or to become the substrate of patents, across time and diverse sub-domains. The choice now is not between subjective, qualitative assessment and simple quantitative metrics, but with machine learning models no less objective, reproducible, and generalisable for being complex. We cannot and do not argue that machine learning models remove the need for qualitative assessment, but only that the quantitative metrics in current use could be far better. Indeed, the clearly observed relation between model complexity and achieved fidelity suggests modelling of the body of a paper—currently infeasible for copyright reasons—is likely to yield still higher fidelity. This will inevitably usher an examination of policies on the right trade-off between performance and intelligibility that must be settled politically, not empirically.

Our models are of direct, first-order inclusions, indifferent to the upstream published sources a given paper itself cites. They may be more likely to predict the translational potential of a meta-analysis, for example, than that of any of the preceding studies informing it. But the proposed framework can be naturally extended to second or higher order inclusions earlier in the citation path, weighting the cascade of information down the full translational pathway in a principled way. For example, the citation nexus has been modelled as a graph (20) with publication-based metrics as the predictive target, in evolution of established approaches for predictive modelling of bibliographically defined impact(21, 22). Note the constraint on inclusion depth amongst other considerations prevents the naïve use of our models to determine the causal *sufficiency* of translation, but no-one would claim that *any* metric within so complex a system could plausibly index causality on its own. A complex model can also be used to distinguish empirical from meta-analytical papers with potentially greater accuracy than bibliographic "article type" tags, weighting inclusions their empirical content.

Equally, while unethical biases can corrupt carelessly designed or interpreted complex models, they can also be revealed by them where the neglected subpopulation is defined by the complex interaction of several variables of ethical concern simple models are too crude to illuminate. Insisting on simple, low-dimensional decision boundaries does not remove bias but merely conceals it from view: complex models—correctly designed and used—are not the problem here but an essential part of the solution. A sharp distinction must be drawn between simplicity and explainability: where a system is inherently, irreducibly complex, a simple metric *cannot* be explanatory. The unprecedented scale of analysed data, drawn from the largest open bibliographic repository in the world, limits potential distortion from sampling bias; the use of out-of-sample, ahead-of-time measures of performance further strengthens generalisability. Any individual or institution can submit test data to the model, and independently validate predictions over time, or retrain with further, prospectively acquired data.

Our work builds on existing research on patent, guideline and policy inclusions. A "patent-paper citation index" has been proposed to formalise science-to-technology linkages (23), and patent inclusions have been systematically evaluated to quantify value return on public research investments (24) and used as a marker of the technological importance of scientific papers (25). While it may seem that patents should precede published research, a large study of US patent and paper linkage found that 60% referenced prior research (26). Patent inclusions have therefore been explored as indicators of papers whose recognition has been delayed. Similarly, a focus on impact assessments has prompted analyses of referencing patterns within cancer guidelines(27), small hand-curated groups of guidelines(28), and separately, policy inclusions extracted by hand(29), systematic analysis of COVID-19 policy(30), or from Altmetric(31) — though difficulties with comprehensive data acquisition have hampered the latter. Although one study has recently attempted to predict combined guideline and clinical trial citations of basic research using a small set of MeSH term derived features(32), no predictive framework for the tangible product of scientific research, rather than trials, has been previously described. The critique of paper citation metrics for measuring impact is not new, but the argument can now be rigorously tested against objective markers of translation.

The application of highly expressive language models to searchable, comprehensive, fully-digitized repositories of scientific publications has the power to derive compact—yet rich—representations of research activity on which high-fidelity predictive models can be founded. Here focused on the task of predicting translational signals, the approach can be used to forecast many aspects of scientific activity upstream of real-world impact. Our work argues for a radical shift towards the adoption of novel methods in the evaluation of medical research, a shift for which observed levels of translational productivity—declining for more than half a century—demand urgent and decisive action.

## MATERIALS AND METHODS

### Data

The dataset was downloaded from Microsoft Academic Graph (MAG) in March 2019(17). It was filtered to include medical papers—as labelled by MAG—published from January 1990 to March 2019, with at least one paper-to-paper citation. To extract a patent inclusion count, papers were matched by id to the reference list on patent entries, in turn provided within MAG through the Lens database(33). To extract guideline or policy inclusion counts, papers were matched by title to a dataset kindly provided by the Wellcome Trust, containing reference lists scraped from documents on the World Health Organisation, National Institute of Clinical Excellence, Unicef, Médecins Sans Frontières, UK Government and UK Parliament websites. A free web-based tool for guideline and policy inclusion detection is available from Wellcome Data Labs (https://reach.wellcomedatalabs.org/) and associated code is available (https://github.com/wellcometrust/reach). Title matching was by a combination of fuzzy matching and cosine similarity of term frequency-inverse document frequency vectors, with manual cleaning of the resulting matches.

The full feature list extracted from MAG is included in Table S2, and summarily comprised publication year; paper citation count; paper rank; author count; reference count; and rank, paper count and paper citation information for the first and last author, the first and last author's affiliations, the journal and the field. The first level of medical domain fields were extracted—43 in total—and added as features using multiple one hot encoding. Field names from hierarchical topic modelling were supplied within MAG(34), and rank — a Reinforcement Learning estimation of dynamic eigencentrality, reflecting a paper's connectedness to other influential entries in the graph(35)—was also supplied within MAG. In addition to a simple paper citation count, the number of times a paper was referenced in the text body of another paper was summed to create a 'paper mentions' count.

### Predictive analysis

NATURAL LANGUAGE PROCESSING. Medical papers were further filtered to include those with titles and abstracts. Sentence level embeddings were generated for each title using BioBERT(18), a state-of-the-art BERT language model pre-trained on biomedical corpora comprising PubMed abstracts and PMC full text articles, in addition to general corpora comprising English Wikipedia and BooksCorpus. BERT is a highly influential Transformer encoder, released in 2018, that is able to learn the context of words by joint conditioning on the full sentence, rather than creating a sequential representation where context is lost with increasing distance between words(36). The sentence level embeddings were derived from the output of the first ([CLS]) token.

To create a fixed length abstract level embedding, we truncated the abstracts to 20 sentences or zero padded where the abstract was shorter, replacing each sentence with its BioBERT embedding and concatenating the array to create a 15360-dimensional vector. This was further concatenated with the title vector, creating a 16128-dimensional representation of the title and abstract taken together.

PREPROCESSING. To rebalance the proportions of positive and negative target labels, the majority negative class was randomly sampled without replacement. Papers without a title or abstract were then removed. This led to a 1.1:1 balance of positive to negative labels in the patent group and the guideline or policy group. Data was randomly split into label stratified training and test sets with a 9:1 ratio. Missing values in the metadata were imputed with medians derived from the training split, and values were transformed into z-scores.

MODELLING. To address the primary objective of detecting signals of translation, we trained a series of models to predict a binary outcome of inclusion in a patent vs none, and separately a binary outcome of inclusion in a guideline or policy document vs none. This was motivated by two considerations, first that each outcome was an independent measure of translation rewarding predominantly basic science or clinical science in patent and guideline or policy classes respectively, and second, that patent inclusions were around 50 times more prevalent than guideline or policy inclusions and might unfairly dwarf the predictive signal of the latter class.

We first modelled a single variable—paper citations per year—using logistic regression, to provide a baseline prediction reflecting current citations-based practice. The hyperparameters of this model were optimised using a parallelised, cross-validated gridsearch. Second, we modelled the metadata—all features extracted from MAG pertaining to the paper and its research environment, excluding title and abstract embeddings— using a 6 layer perceptron, with categorial variables one-hot encoded. Third, we trained a 1-dimensional CNN on the title and abstract embeddings, using an initial kernel length and stride of 768 to match the length of each sentence vector. We employed the common architectural heuristics of iteratively decreasing layer sizes and iteratively increasing the number of output channels to a final set of fully connected layers. These two models were tuned by cross-validation within the training set, and the best models were combined into a final model that took both the metadata and title-abstract embeddings as inputs, specified in Fig. S1.

INTERPRETABILITY. Deep neural networks do not explicitly provide quantification of the importance of individual features to prediction. We therefore trained and grid-search optimised an AdaBoost model(37) on metadata features and extracted Gini-importance from the best performing model on validation data.

Further to illuminate the title-abstract embeddings, we trained a fully-connected autoencoder on the 16128 BioBERT dimensions of each title and abstract, deriving a 50 dimensional representation further compressed to two dimensions with t-Distributed Stochastic Neighbour Embedding (t-SNE)(38). The resulting plot was coloured by the presence or absence of a translation inclusion.

MODEL EVALUATION. The predictive performance of all models on the training set was evaluated by stratified 10-fold cross-validation using area under the receiver operating curve (AUROC) and average precision (AP), a measure of the area under the precision recall curve. The former is a common metric for assessing predictive performance which balances sensitivity against specificity across a range of classification thresholds, the latter is more resistant to imbalanced data bias and balances sensitivity against precision, the purity of predicted positive results. The final, tuned, highest performing model was tested on the unseen test data and assessed by AUROC and AP.

To assess the performance of the final model on future papers, the same architecture was trained from scratch on data from Jan 1990 to Dec 2013, and tested on data from Jan 2014 to Dec 2017. We measured the performance in each of the four years, and within the top 8 fields to investigate the calibration to these groups. To quantify any reliance on time-dependent citation patterns for a given paper, we assessed the performance of the full model whose training set had 'paper citation count', 'paper rank' and 'paper mentions count' variables removed; similarly, to quantify any reliance on features denoting merit extrinsic to the paper, author, institution, journal and field –level ranks and counts were removed.

As further validation, an external, publically available dataset containing the publication output of Nobel Prize Laureates in Physiology or Medicine(39) was downloaded, matched to MAG, and processed identically to the test data. All papers from 1991-2019 published prior to—and including—the prize-winning paper were tested for patent inclusion prediction using: a logistic regression model of age-normalised-citations; the MLP of extended metadata; and the hybrid model of extended metadata and title-abstract embeddings, each pre-trained on the entire training corpus, with the Nobel papers removed. The AUCs and numbers assigned to positive and negative labels were recorded.

## Descriptive analysis

As a secondary objective, we sought to understand the correspondence of patent and guideline or policy citations to the far more widely measured and acknowledged paper citations, as well as to understand the community structure of patent included vs non-included groups, and guideline or policy included vs not included groups.

Towards the former aim, we plotted paper citations against translation inclusions, and examined their correlation by fitting a linear regression model with 1000x bootstrapped confidence intervals. We further ranked journals by paper citation counts normalised by the journal's total paper count within our dataset—filtered as described to medical papers from 1990-2019 with at least one citation. This roughly corresponds to a canonical 'Impact Factor', though the interval is widened from yearly to three decades. We repeated this for a journal's patent inclusions count and guideline or policy inclusion count. Journals analysed in this manner were filtered to include only those with 500 or more total papers in the dataset.

Towards the latter aim, we fit Bayesian weighted, non-parametric, nested stochastic block models(19) on all papers with patent inclusions, and all papers without them, and then again on all papers with guideline or policy inclusions, and all papers without them, degree-corrected and weighted exponentially by the absolute value of the pairwise correlations of features extracted from MAG (excluding titles and abstracts). Stochastic block models are generative random graph models which display community structures, subsets of nodes connected by larger edge densities than those outside the subset. The models were strengthened by sampling from the posterior distribution and equilibrated with Markov Chain Monte Carlo over 100,000 iterations to ensure convergence. Scalable force directed placement(40) was used for visualisation of the combined feature graph, with node size proportional to eigencentrality, and edge weight and colour proportional to the absolute value of the correlation coefficient between two features.

ANALYTIC ENVIRONMENT. All analyses were written in Python 3.5. Preprocessing was performed using Pandas(41), NumPy(42) and Scikit-Learn(43), and visualisation using Matplotlib(44), Seaborn(45), Graph-tool(46). Neural networks were built in Keras(47) with Tensorflow backend, and PyTorch(48), other models were built in Scikit-Learn. t-SNE was performed using Multicore-TSNE(49), BioBERT models were downloaded and implemented locally. The hardware specification used was: 96 GB RAM, Intel® Xeon(R) CPU E5-2620 v4 @ 2.10 GHz × 32 processor, and GeForce GTX 1080/PCIe/SSE2 graphics.

# References

1. N. D. Bellis, *Bibliometrics and Citation Analysis: From the Science Citation Index to Cybermetrics* (Scarecrow Press, 2009).

2. E. Garfield, A Unified Index to Science. *Proceedings of the International Conference on Scientfic Information*, **1**, 461–74 (1968).

3. D. J. Price, *Little science, big science-- and beyond* (Columbia University Press, 1986).

4. J. M. Nicholson, J. P. A. Ioannidis, Conform and be funded. *Nature* **492**, 34–36 (2012).

5. G. Lewison, R. Cottrell, D. Dixon, Bibliometric indicators to assist the peer review process in grant decisions. *Res Eval* **8**, 47–52 (1999).

6. V. M. Patel, *et al.*, How has healthcare research performance been assessed?: a systematic review. *J R Soc Med* **104**, 251–261 (2011).

7. K. El Emam, L. Arbuckle, E. Jonker, K. Anderson, Two h-Index Benchmarks for Evaluating the Publication Performance of Medical Informatics Researchers. *J Med Internet Res* **14** (2012).

8. L. L. Haak, *et al.*, The electronic Scientific Portfolio Assistant: Integrating scientific knowledge databases to support program impact assessment. *Sci Public Policy* **39**, 464–475 (2012).

9. A. Angeli, *et al.*, Genotype and phenotype in Parkinson's disease: Lessons in heterogeneity from deep brain stimulation. *Movement Disorders* **28**, 1370–1375 (2013).

10. M. McNutt, The measure of research merit. *Science* **346**, 1155–1155 (2014).

11. J. E. Hirsch, Superconductivity, what the H? The emperor has no clothes. *arXiv:2001.09496 [cond-mat, physics:physics]* (2020) (September 9, 2021).

12. P. Nachev, D. Herron, N. McNally, G. Rees, B. Williams, Redefining the research hospital. *npj Digital Medicine* **2**, 1–5 (2019).

13. D. G. Contopoulos-Ioannidis, E. Ntzani, J. P. A. Ioannidis, Translation of highly promising basic science research into clinical applications. *Am. J. Med.* **114**, 477–484 (2003).

14. N. Graves, A. G. Barnett, P. Clarke, Cutting random funding decisions. *Nature* **469**, 299–299 (2011).

15. M. Liu, *et al.*, The acceptability of using a lottery to allocate research funding: a survey of applicants. *Research Integrity and Peer Review* **5**, 3 (2020).

16. S. Avin, Mavericks and lotteries. *Studies in History and Philosophy of Science Part A* **76**, 13–23 (2019).

17. A. Sinha, *et al.*, An Overview of Microsoft Academic Service (MAS) and Applications in *Proceedings of the 24th International Conference on World Wide Web*, WWW '15 Companion., (Association for Computing Machinery, 2015), pp. 243–246.

18. J. Lee, *et al.*, BioBERT: a pre-trained biomedical language representation model for biomedical text mining. *Bioinformatics* **36**, 1234–1240 (2020).

19. T. P. Peixoto, Hierarchical Block Structures and High-Resolution Model Selection in Large Networks. *Phys. Rev. X* **4**, 011047 (2014).

20. J. W. Weis, J. M. Jacobson, Learning on knowledge graph dynamics provides an early warning of impactful research. *Nature Biotechnology*, 1–8 (2021).

21. D. E. Acuna, S. Allesina, K. P. Kording, Predicting scientific success. *Nature* **489**, 201–202 (2012).

22. L. Fu, C. Aliferis, Using content-based and bibliometric features for machine learning models to predict citation counts in the biomedical literature. *Scientometrics* **85**, 257–270 (2010).

23. Y. Yamashita, Exploring Characteristics of Patent-Paper Citations and Development of New Indicators. *Scientometrics* (2018) https:/doi.org/10.5772/intechopen.77130 (June 7, 2020).

24. D. Li, P. Azoulay, B. N. Sampat, The applied value of public investments in biomedical research. *Science* **356**, 78–81 (2017).

25. A. F. J. van Raan, Patent Citations Analysis and Its Value in Research Evaluation: A Review and a New Approach to Map Technology-relevant Research. *Journal of Data and Information Science* **2**, 13–50 (2017).

26. M. Ahmadpoor, B. F. Jones, The dual frontier: Patented inventions and prior scientific advance. *Science* **357**, 583–587 (2017).

27. G. Lewison, R. Sullivan, The impact of cancer research: how publications influence UK cancer clinical guidelines. *British Journal of Cancer* **98**, 1944–1950 (2008).

28. J. Grant, R. Cottrell, F. Cluzeau, G. Fawcett, Evaluating "payback" on biomedical research from papers cited in clinical guidelines: applied bibliometric study. *BMJ* **320**, 1107–1111 (2000).

29. R. Newson, L. Rychetnik, L. King, A. Milat, A. Bauman, Does citation matter? Research citation in policy documents as an indicator of research impact – an Australian obesity policy case-study. *Health Research Policy and Systems* **16**, 55 (2018).

30. Y. Yin, J. Gao, B. F. Jones, D. Wang, Coevolution of policy and science during the pandemic. *Science* **371**, 128–130 (2021).

31. R. Haunschild, L. Bornmann, How many scientific papers are mentioned in policy-related documents? An empirical investigation using Web of Science and Altmetric data. *Scientometrics* **110**, 1209–1216 (2017).

32. B. I. Hutchins, M. T. Davis, R. A. Meseroll, G. M. Santangelo, Predicting translational progress in biomedical research. *PLoS Biol* **17**, e3000416 (2019).

33. O. A. Jefferson, *et al.*, Mapping the global influence of published research on industry and innovation. *Nature Biotechnology* **36**, 31–39 (2018).

34. Z. Shen, H. Ma, K. Wang, A Web-scale system for scientific knowledge exploration. *arXiv:1805.12216 [cs]* (2018) (June 2, 2020).

35. K. Wang, *et al.*, A Review of Microsoft Academic Services for Science of Science Studies. *Front. Big Data* **2** (2019).

36. J. Devlin, M.-W. Chang, K. Lee, K. Toutanova, BERT: Pre-training of Deep Bidirectional Transformers for Language Understanding. *arXiv:1810.04805 [cs]* (2019) (June 2, 2020).

37. Y. Freund, R. E. Schapire, A Decision-Theoretic Generalization of On-Line Learning and an Application to Boosting. *Journal of Computer and System Sciences* **55**, 119–139 (1997).

38. L. van der Maaten, G. Hinton, Visualizing Data using t-SNE. *Journal of Machine Learning Research* **9**, 2579–2605 (2008).

39. J. Li, Y. Yin, S. Fortunato, D. Wang, A dataset of publication records for Nobel laureates. *Scientific Data* **6**, 33 (2019).

40. Y. Hu, Efficient, High-Quality Force-Directed Graph Drawing. *The Mathematica Journal*, 35 (2006).

41. W. McKinney, Data Structures for Statistical Computing in Python in (2010), pp. 51–56.

42. T. E. Oliphant, Python for Scientific Computing. *Computing in Science and Engg.* **9**, 10–20 (2007).

43. F. Pedregosa, *et al.*, Scikit-learn: Machine Learning in Python. *Journal of Machine Learning Research* **12**, 2825–2830 (2011).

44. J. D. Hunter, Matplotlib: A 2D Graphics Environment. *Computing in Science Engineering* **9**, 90–95 (2007).

45. Michael Waskom, *et al.*, seaborn: v0.5.0 (2014).https:/doi.org/10.5281/zenodo.12710 (June 11, 2020).

46. T. P. Peixoto, *The graph-tool python library* (2014).

47. F. Chollet, and others, *Keras* (2015).

48. A. Paszke, *et al.*, Automatic differentiation in PyTorch (2017) (January 20, 2021).

49. D. Ulyanov, Multicore-tsne. *GitHub Repos. GitHub* (2016).

# Acknowledgements

## Funding

APKN, RG, PN, DH, NM & BW are supported by the NIHR UCLH Biomedical Research Centre. PN is supported by the Wellcome Trust.

## Author contributions

APKN, PN, DH, NM and BW conceived the study; NS and DM provided data on guideline and policy references; JKR performed graphical descriptive analysis; HCW contributed to the natural language processing method; RJG contributed to the design and evaluation of deep learning models. PN directed the study and contributed to draft typescript. APKN designed, coded and evaluated the study and drafted the typescript. All authors provided critical editorial support to the typescript.

## Competing Interests

The authors declare that they have no competing interests.

## Data and materials availability

The data that support the findings of this study are available on application to Microsoft Academic Graph at https://www.microsoft.com/en-us/research/project/microsoft-academic-graph/. Guideline and policy data are available from the Reach project at Wellcome Data Labs https://reach.wellcomedatalabs.org/. Code for extracting guideline and policy references is available at https://github.com/wellcometrust/reach. Analytic code will be made available upon reasonable request.

# Figures and Tables

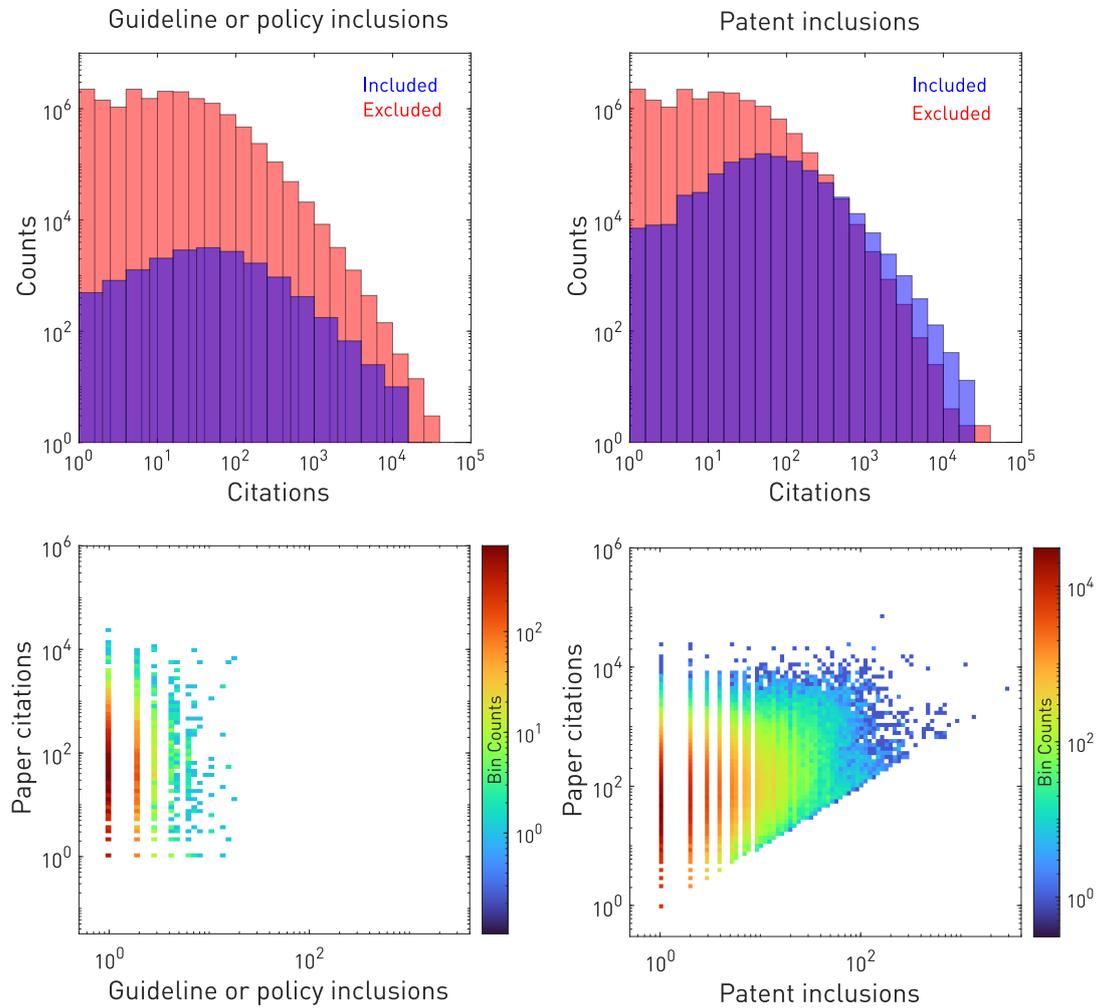

Fig. 1. Citation histograms of papers included (red) or not included (blue) in guideline or policy documents (top left) or patents (top right), plotted on log axes. The area of overlap is shown in purple, and the contrasting papers are drawn from a random sample of the same size as each included group. The relationship between citation and inclusion counts for included papers is shown in binned scatterplots for guideline or policy inclusions (bottom left) and patent inclusions (bottom right), also plotted on log axes.

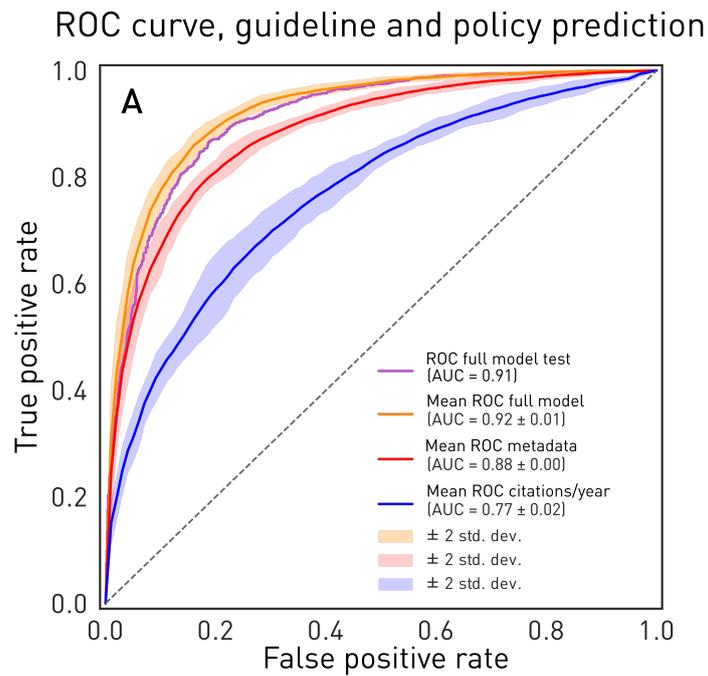

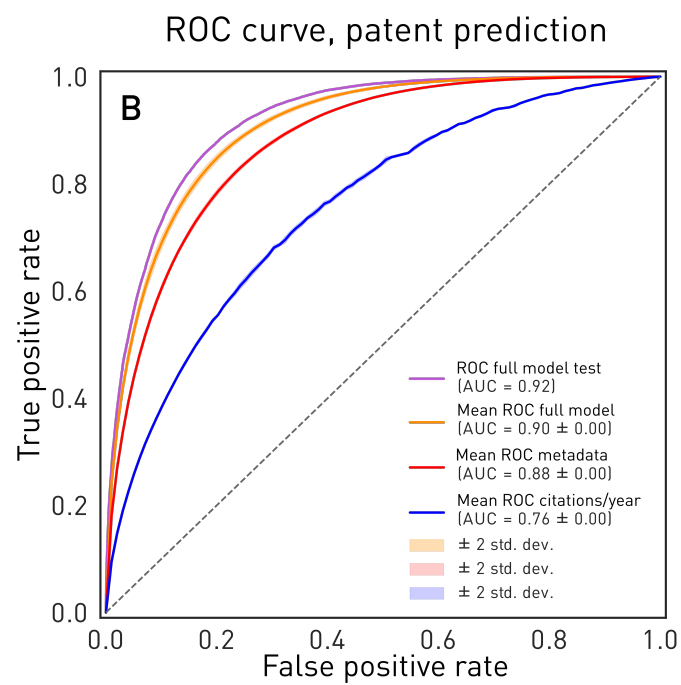

Fig. 2. Model predictive performance. Cross-validated ROC curves for the hybrid CNN and MLP model trained on metadata and title and abstract embeddings (orange for validation, purple for held-out test); the MLP model trained on metadata (red), and the logistic regression model trained on citation count per year (blue) for guideline or policy inclusions (A) and patent inclusions (B). The confidence intervals are ± 2 standard deviations on cross-validation.

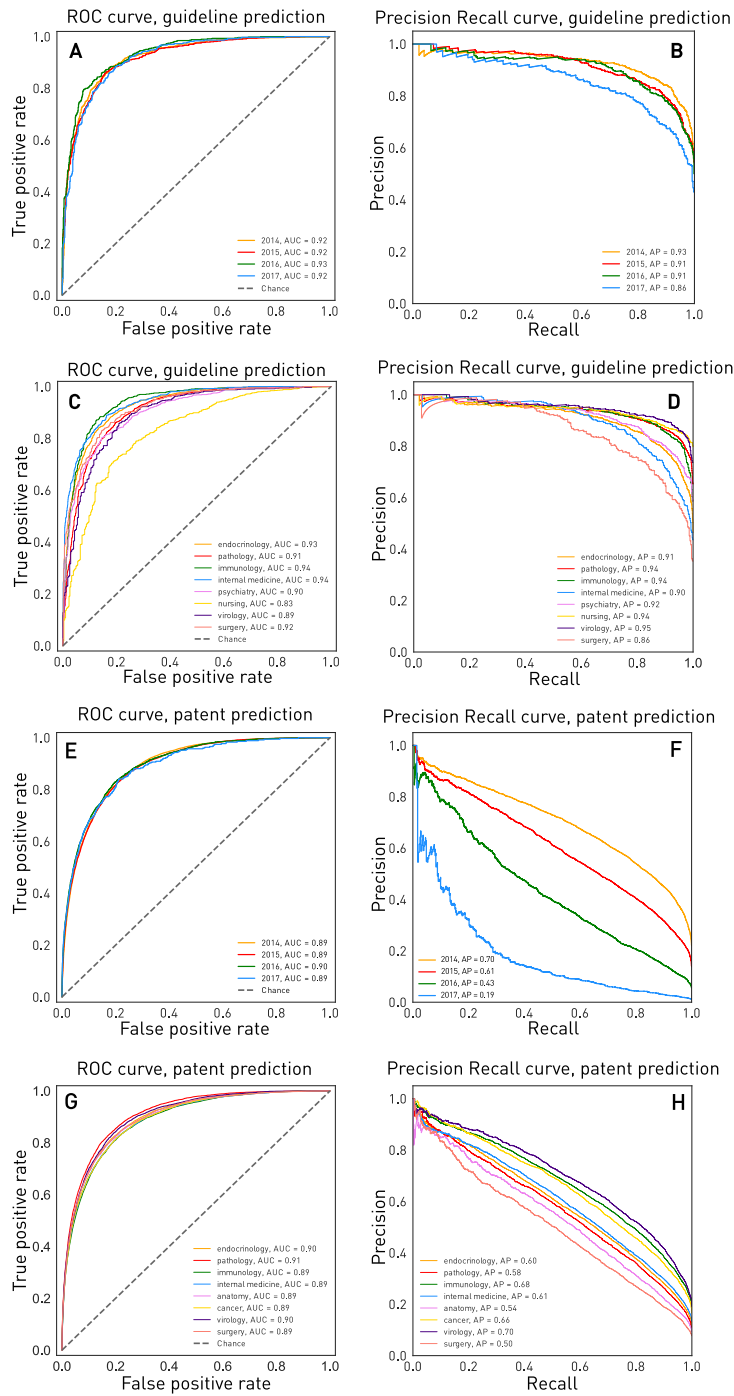

**Fig. 3.** ROC curves (A, C), and Precision-Recall curves (B, D) for the hybrid CNN and MLP model trained on data from 1990 to 2013, and tested on papers published in the subsequent 4 years, plotted by year, for guideline or policy inclusions and patent inclusions respectively. ROC curves (E, G), and Precision-Recall curves (F, H) for combination CNN and MLP model trained on data from 1990-2013, and tested on data from 2014-2019, plotted by each of the top 8 most common fields, for guideline or policy inclusions and patent inclusions respectively.

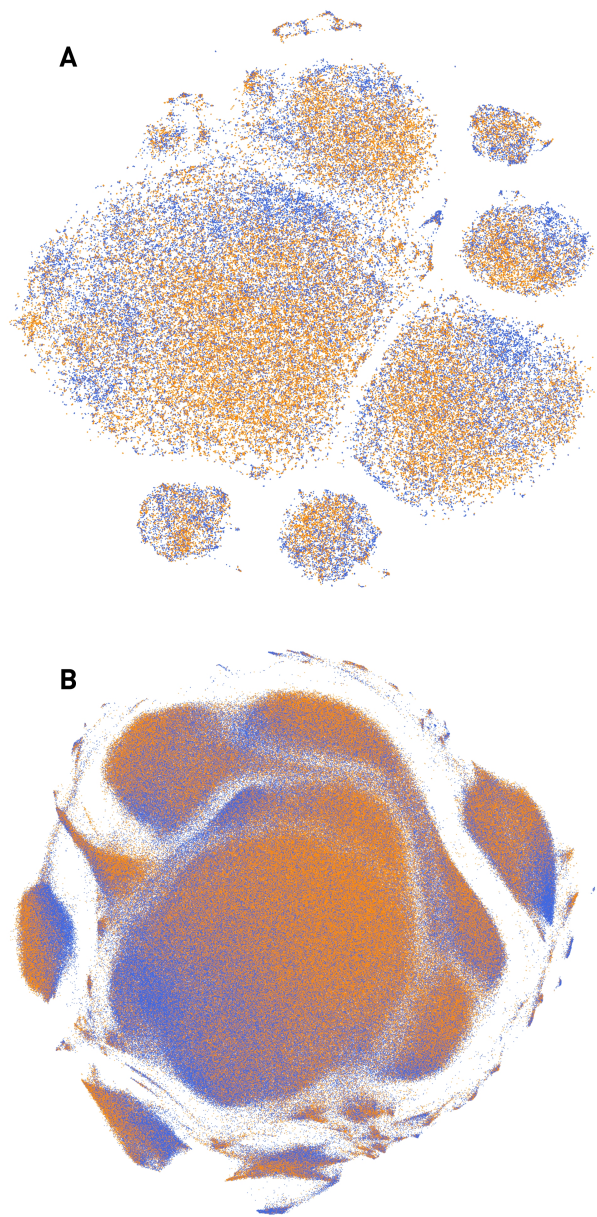

Fig. 4. t-Distributed Stochastic Neighbour Embedding (t-SNE) projection in 2-dimensions of title and abstract BioBERT autoencoder embeddings. Labelled by the presence (orange) or absence (blue) of a guideline or policy (A) or patent (B) inclusion. Note discernible data structure that enables accurate prediction of inclusion, though too complex to be reduced to any small set of characteristic features.

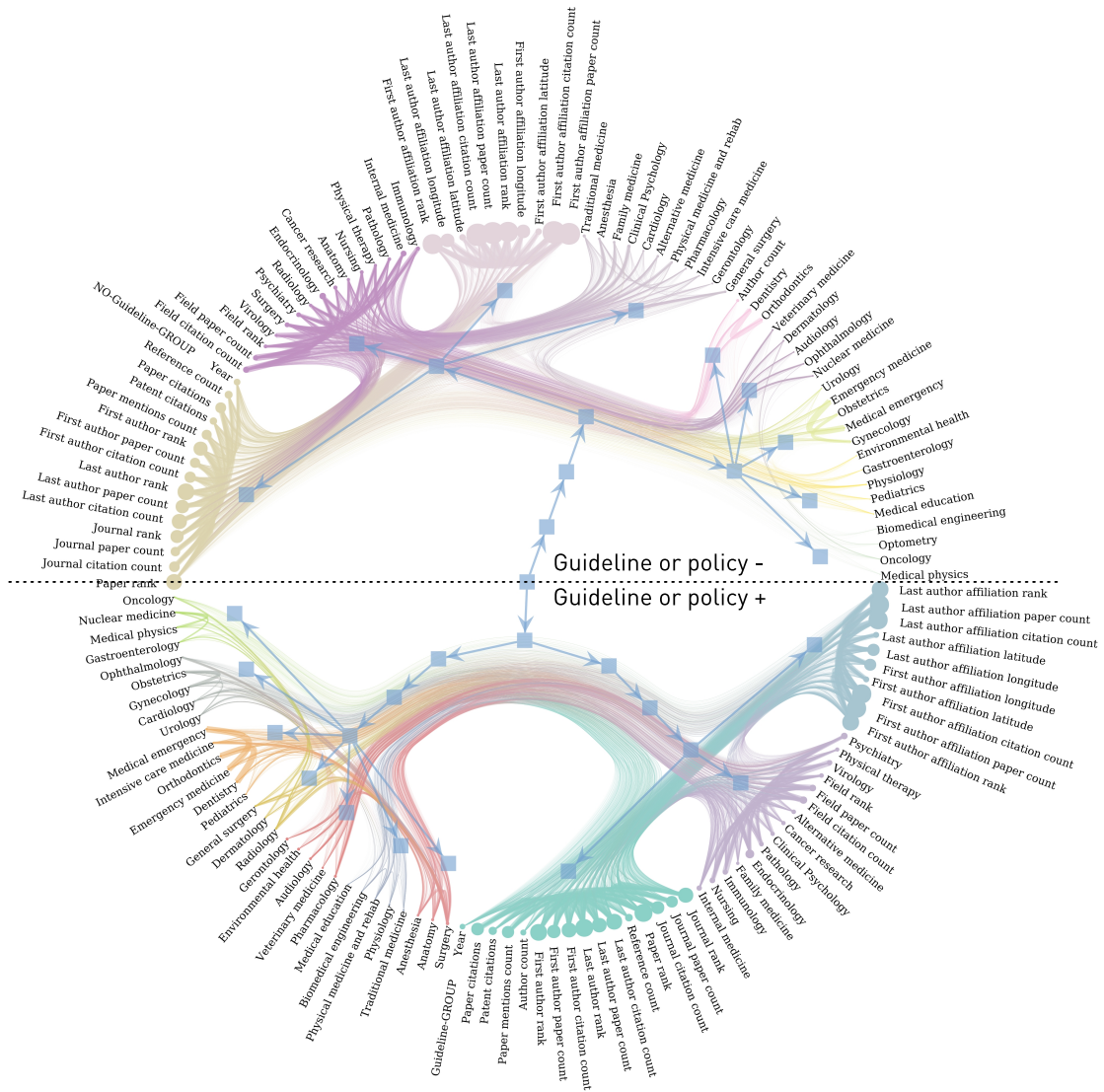

Fig. 5. Nested stochastic block models (SBM) showing the community structure of the metadata of papers included in guidelines or policy (top) vs those not (bottom). Node size in these models corresponds to the eigencentrality of each feature, and edge weight corresponds to pairwise absolute value of correlation coefficient between features. The included class is the bottom hemifield.

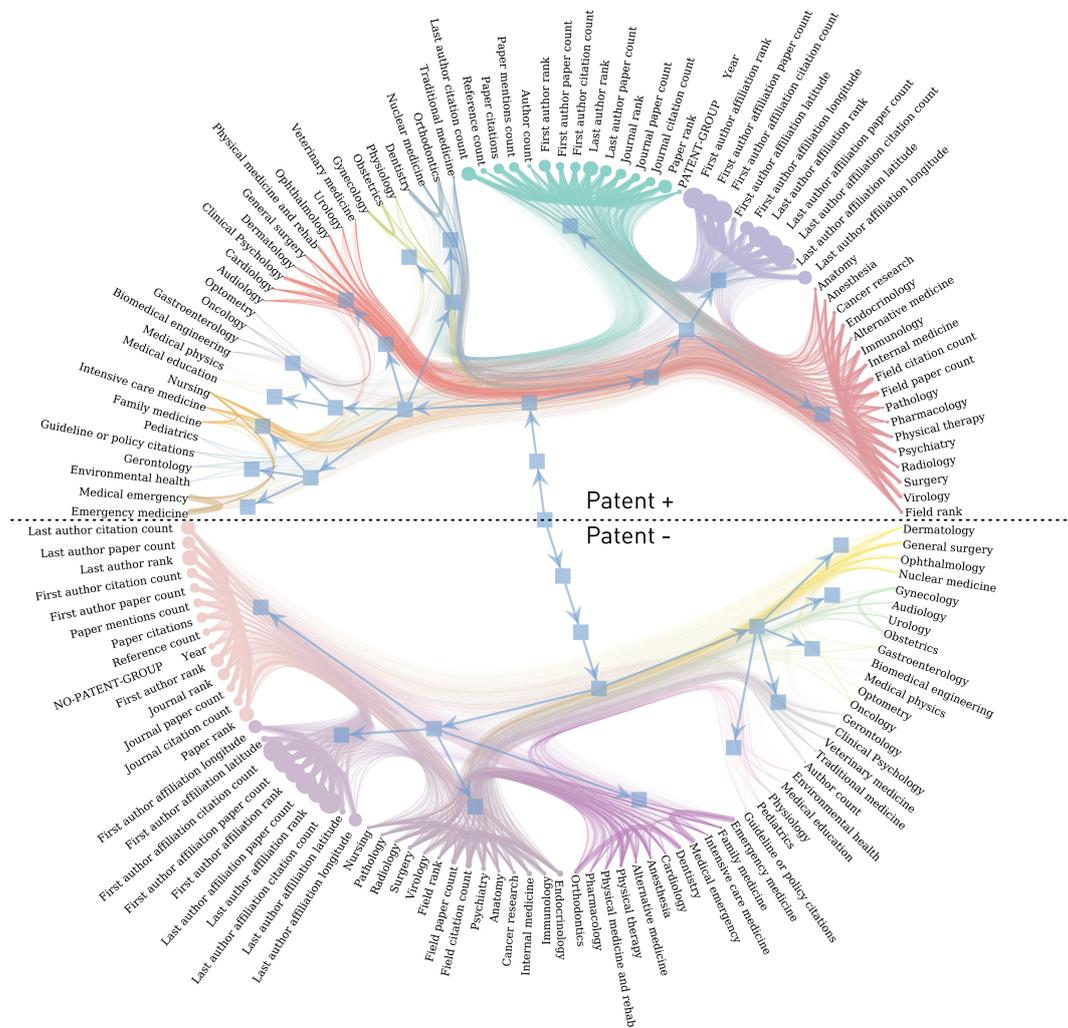

Fig. 6. Nested stochastic block models (SBM) showing the community structure of the metadata of papers included in patents (top) vs those not (bottom). Node size in these models corresponds to the eigencentrality of each feature, and edge weight corresponds to pairwise absolute value of correlation coefficient between features. The included class is the top hemifield.

Table 1A. Top 10 journals by paper citations per paper

| Journal | Paper citations / total papers | Paper citations | Total papers |
| --- | --- | --- | --- |
| Annual Review of Immunology | 480.835 | 376494 | 783 |
| Physiological Reviews | 407.457 | 392789 | 964 |
| Annual Review of Neuroscience | 383.877 | 240691 | 627 |
| Psychological Bulletin | 333.302 | 409295 | 1228 |
| Pharmacological Reviews | 318.500 | 210847 | 662 |
| Cell | 295.994 | 3229298 | 10910 |
| Annual Review of Psychology | 283.465 | 161575 | 570 |
| CA: A Cancer Journal for Clinicians | 263.673 | 240733 | 913 |
| Psychological Review | 262.322 | 247107 | 942 |
| Clinical Microbiology Reviews | 250.334 | 239319 | 956 |

Table 1B. Top 10 journals by guideline or policy inclusions per paper

| Journal | Guideline or policy inclusions / total papers | Guideline or policy inclusions | Total papers |
| --- | --- | --- | --- |
| Tobacco Control | 0.094 | 3120 | 3325 |
| Eastern Mediterranean Health Journal | 0.091 | 282 | 3090 |
| Noise & Health | 0.085 | 55 | 649 |
| Human Resources for Health | 0.084 | 69 | 824 |
| Health Policy and Planning | 0.079 | 157 | 1977 |
| Influenza and Other Respiratory Viruses | 0.061 | 63 | 1036 |
| Globalization and Health | 0.043 | 29 | 681 |
| PLOS Medicine | 0.040 | 152 | 3809 |
| Bulletin of The World Health Organization | 0.039 | 247 | 6278 |
| Trauma, Violence, & Abuse | 0.038 | 22 | 574 |

Table 1C. Top 10 journals by patent inclusions per paper

| Journal | Patent inclusions / total papers | Patent inclusions | Total papers |
| --- | --- | --- | --- |
| Annual Review of Immunology | 5.733 | 4489 | 783 |
| Nature Biotechnology | 3.862 | 17042 | 4413 |
| Protein Engineering | 3.153 | 2557 | 811 |
| Pharmacological Reviews | 3.045 | 2016 | 662 |
| Trends in Biotechnology | 2.990 | 4733 | 1583 |
| Cell | 2.725 | 29729 | 10910 |
| Journal of Experimental Medicine | 2.293 | 24105 | 10513 |
| Advanced Drug Delivery Reviews | 2.167 | 6862 | 3166 |
| Chemical Reviews | 1.925 | 1305 | 678 |
| Transfusion Science | 1.845 | 1552 | 841 |